\documentclass[groupedaddress,twocolumn,prl,letterpaper]{revtex4}
\usepackage{epsfig,amsfonts,amsmath}
\newcommand{\be}{\begin{equation}}
\newcommand{\ee}{\end{equation}}
\newcommand{\ba}{\begin{eqnarray}}
\newcommand{\ea}{\end{eqnarray}}

\def\bea{\begin{eqnarray}}
\def\eea{\end{eqnarray}}
\def\ben{\begin{eqnarray*}}
\def\een{\end{eqnarray*}}

\def\>{\rangle}
\def\<{\langle}
\def\l{\left}
\def\r{\right}
\newcommand{\nc}{\newcommand}
\nc{\cN}{{\cal N}}

\def\l{\lambda}

\def\r{\rho}

\def\be{\begin{equation}}
\def\ee{\end{equation}}
\def\bea{\begin{eqnarray}}
\def\eea{\end{eqnarray}}
\def\ben{\begin{eqnarray*}}
\def\een{\end{eqnarray*}}

\def\>{\rangle}
\def\<{\langle}

\def\l{\left}
\def\r{\right}

\newcommand{\fig}[1]{Fig.~\ref{fig:#1}}

\newcommand{\ord}[1]{\mathcal{O}(#1)}

\begin{document}

\title{Nanofabrication by magnetic focusing of supersonic beams}
\author{Robert J. Clark, Thomas R. Mazur, Adam Libson, and Mark G. Raizen}
\affiliation{Center for Nonlinear Dynamics and Department of Physics, The University of Texas at Austin, Austin, TX, 78712, USA}

\begin{abstract}

We present a new method for nanoscale atom lithography. We propose the use of a supersonic atomic beam, which provides an extremely high-brightness and cold source of fast atoms. The atoms are to be focused onto a substrate using a thin magnetic film, into which apertures with widths on the order of 100~nm have been etched. Focused spot sizes near or below 10~nm, with focal lengths on the order of 10~$\mu$m, are predicted. This scheme is applicable both to precision patterning of surfaces with metastable atomic beams and to direct deposition of material. 

\end{abstract}

\pacs{} 

\maketitle

Nanoscale fabrication is a critical tool for realizing much of modern technology, including information processing, biomedical research, and photonics \cite{microfab:book,Obrien:09,Dubertret:02}. Optical lithography, the current method for chip mass production, is used to produce many features in parallel, but has a limited resolution due to the wavelength of light used. Electron beam (e-beam) lithography, a method for producing much smaller (order of 1~nm) features, is a serial, rather than parallel, method, meaning that it is much more time-consuming than optical lithography. E-beam, however, is frequently used to fabricate the masks that are required by the latter. Additional methods, using vacuum ultraviolet or X-ray radiation \cite{Chen:01}, or using focused ion beams \cite{Orloff:93}, are being developed in an effort to achieve high-resolution and high-throughput nanofabrication. 

One area of great potential for nanofabrication is atom lithography \cite{Meschede:03,Balykin:09b}. The deBroglie wavelength of atoms is typically much less than an optical wavelength, potentially resulting in a much smaller diffraction-limited spot size. Additionally, fabrication operations are parallelizable; much work has focused on depositing multiple lines and dots of atoms by focusing from a standing light wave  \cite{Johnson:98,Baker:04,Lu:98,Bard:97,Engels:99}. In addition, atom lithography is versatile in the sense that one may either directly write structures onto a substrate  \cite{McClelland:93,Timp:92,Natarajan:96,Gupta:95,Smeets:09}, or may pattern a resist prior to etching, as in traditional lithography, using a beam of metastable atoms \cite{Berggren:95}. The primary limitation of optical focusing is that it is difficult to focus some of the atoms without simultaneously defocusing others, leading to significant aberrations and an unwanted underlayer of material. A method of mitigating this effect was proposed \cite{Averbukh:01} and demonstrated \cite{Oskay:02}, but is challenging to implement in practice. Alternative approaches, such as focusing of atoms using macroscopic magnetic lenses \cite{Kaenders:95} or an ``atom pinhole camera'' \cite{Balykin:09c} have also been explored. All the above techniques have used effusive beams, limiting atomic density to around $10^{10}$ atoms/cm$^3$, and have achieved controlled feature sizes (at best) on the order of 100~nm. 

In this Letter, we propose a new approach to atom lithography that should enable much smaller feature sizes and larger throughput. Our method consists of magnetic focusing of a supersonic beam through a nanofabricated magnetic mask. As nearly all atoms are paramagnetic in either the ground state or an accessible metastable state, magnetic focusing is a very general approach. Furthermore, the supersonic beam provides both high atomic flux and a low temperature $T \approx 100$~mK, a unique combination. This paper is organized as follows. We first present details of our proposed method, including the supersonic source and focusing apparatus. We then provide simulation results, including spot sizes and focal lengths for a number of atomic species. Finally, we discuss the scalability of this scheme. 

\begin{figure*}
\begin{center}
\includegraphics[width=\textwidth]{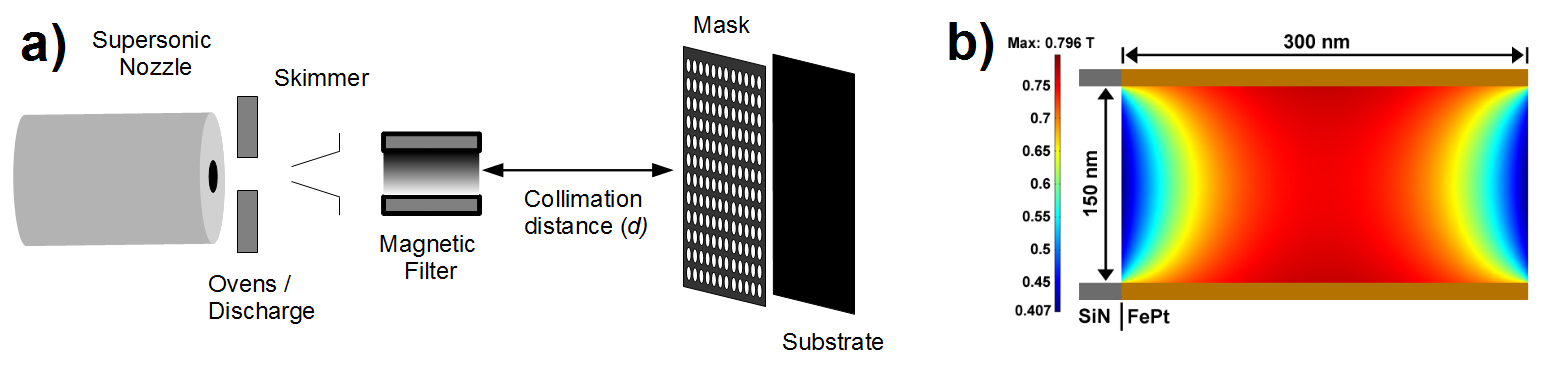} 
\end{center} 
\caption{(a) Schematic (not to scale) of our nanolithography process. The supersonic nozzle produces a bright atomic beam, which may either be excited to a metastable level in a discharge or have atoms entrained into it from some number of ovens. Following this, the beam is collimated by passing through a skimmer. Magnetic filtering ensures that only the correct $m_J$ state arrives at the mask, which then focuses the atoms onto a substrate. (b) Schematic (not to scale) of a single hole in the mask with proposed dimensions given. A Si$_3$N$_4$ substrate supports a 300~nm-thick FePt mask with perpendicular magnetization of 670~kA/m. Only small regions of the mask and substrate are shown; the actual length of the substrate is on the order of 1~$\mu$m. The magnitude of the magnetic field within the hole in the mask is plotted (color online).}
\label{fig:schematic}
\end{figure*}

Our proposed apparatus (see \fig{schematic}) consists of two main components: a supersonic beam of spin-polarized atoms, and a thin magnetic mask into which an array of holes of $\ord{100\rm{~nm}}$ width is etched, through which the atoms are focused onto a substrate. Depending on whether one wishes to deposit material or pattern using excited-state atoms, one either entrains atoms into the beam from one or more ovens, or excites the carrier gas to a metastable level in a discharge. The basic principle of our method is that the very high magnetic field gradients due to the tiny holes in the magnetic material will be able to focus atoms, even though they are travelling at hundreds of meters per second. 

A continuous-wave supersonic beam \cite{Campargue:01,Pauly:00} will provide an atomic flux on the order of $10^{20}$~atoms/sr/s \cite{Buckland:thesis}. A typical  fraction of either metastable or entrained atoms in the beam is $10^{-3}$. Following a skimmer, atoms in a specific internal magnetic sublevel $m_J$ will be selected by some method of magnetic filtering; either magnetic guiding or deflection may be used. Alternatively, optical pumping could be implemented on many atomic species. For some species, especially metastable noble gases, laser collimation (by transverse laser cooling) may be applied to increase the beam brightness by a factor of $10^3$ or higher, while reducing the velocity spread in the radial direction  \cite{Hoogerland:94,McClelland:93,Rasel:99}. 

The focusing mask consists of a thin film of a magnetized material, deposited on a substrate, with the magnetization vector pointing out of the plane of the film. Such a film (specifically, an FePt film) was recently used to build a permanent-magnet chip trap for atoms \cite{Xing:07,Fernholz:08}. Holes of diameter on the order of 100~nm will be etched into the film and into the substrate that supports it (\fig{schematic}). This substrate could be made of one of many materials, such as silicon nitride (Si$_3$N$_4$). The holes in both layers may be fabricated by conventional e-beam lithography, or possibly by optical lithography. Although the e-beam process is time-consuming, it only needs to be done once to create a ``master'' mask that can be used many times, as in the case of optical lithography. 

The atoms are focused by the force $\vec{F}$ due to the interaction of the atomic magnetic dipole moment $\vec{\mu}$ with the magnetic field $\vec{B}$ of the magnetized mask: $\vec{F} = \nabla \left ( \vec{\mu} \cdot \vec{B} \right )$. Assuming the atomic magnetic dipole adiabatically follows the magnetic field, we may write the force in the radial direction (normal to the propagation direction) as $F_r = -\mu_B g_J m_J \left ( {\partial |B|}/{\partial r} \right )$, where $\mu_B$ is the Bohr magneton and $g_J$ is the Land\'{e} g-factor. For the simulations that follow, we assume a material thickness of 300~nm and a perpendicular magnetization of M~=~670~kA/m  \cite{Xing:07,Fernholz:08}. We also choose a hole diameter of 150~nm. The magnetic fields are computed numerically; the peak field near such a hole is $|B| \approx 0.8$~T (see \fig{schematic}). 

To estimate the number of atoms that passes through each hole per unit time, as well as to estimate the spread in radial velocities, we use a simple model of geometric collimation. The efficiency $\epsilon$ represents the fraction of atoms emitted from a skimmer of radius $r_s$ that will pass through the focusing aperture; it is calculated as the ratio of the area of the circular hole in the mask to the area of the atomic beam at the position of the mask. Writing the beam divergence angle as $\theta$, the distance from the skimmer to the mask as $d$, and the radius of the hole as $r_m$, $\epsilon = r_m^2 / \left ( r_s + d \tan \theta \right)^2$. Typical numbers for our chosen conditions are $r_s = 125$~$\mu$m, $r_m = 75$~nm, $\theta = 7^{\circ}$, and $d = 2$~m, leading to $\epsilon \approx 1 \times 10^{-13}$. Given the beam brightness of $10^{20}$ atoms/sr/s, a discharge or entrainment efficiency of $10^{-3}$, and a loss of one in ten atoms due to magnetic filtering, we estimate a flux through each hole of $\approx 10^{3}$ atoms/s. 

\begin{figure}
\begin{center}
\includegraphics[width=.5\textwidth]{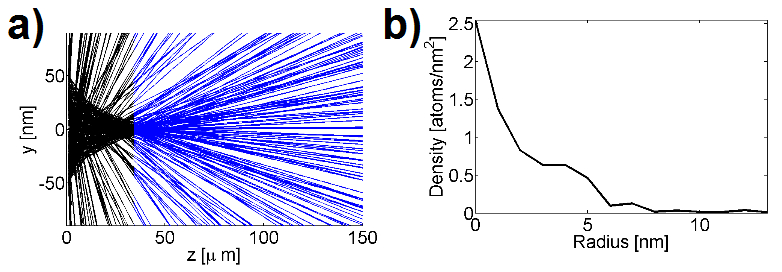}
\end{center} 
\caption{(a) Cross-section of 250 trajectories for $^{20}$Ne with a center-of-mass speed of 400~m/s and a collimation distance of $d = 2$~m. The front face of the mask is located on $z = 0$, and the center of the hole lies on $r = 0$. The spot size $w_0$ is 8.5~nm and the focal length is $f = 34$~$\mu$m. The color of the trajectories changes from black to blue at the focal plane. This spot size is less than the diffraction-limited spot size $w_d = 13.2$~nm; the true spot size will be the result of the convolution of an Airy disc of diameter $w_d$ with the distribution of atoms found via ray tracing. (b) Density of atoms as a function of radius at the focal point for the data in (a).}
\label{fig:truncate}
\end{figure}

The primary goal of our simulations is to calculate the focused spot sizes as the parameters of the problem, including the atomic species and the amount of collimation, are varied. Unless stated otherwise, we assume that the beam is moving with a center-of-mass speed of $v_0 = 400$~m/s, an appropriate value for a neon beam at 77~K. We also assume the radius of the (circular) skimmer is $r_s = 125$~$\mu$m. We assign each atom a random radial velocity, due to geometric collimation, within the range $\Delta v_r \approx 2 r_s v_0/d$, where $d$ is the distance from the skimmer to the mask. The spread in velocities along the propagation axis $\Delta v_z$ is a property of the beam that is due to the supersonic expansion (not to collimation). It is assumed, based on measurements in our laboratory, to be fixed at $\Delta v_z = 14$~m/s for neon at 77~K. Simulations are performed by numerically integrating the equations of motion for a particle moving through the hole in the mask. The spot size $w_0$ is calculated as twice the average value $\langle r \rangle$ of the radius of atoms at the focal point, weighted by the density of atoms at a given radius: $\langle r \rangle = \int \l( \rho(r) \, r \, dr \r)/ \int \l (  \rho(r) \, dr \r )$, where $\rho(r)$ is the density of atoms at radius $r$ and the integral is evaluated  discretely using $dr = 1$~nm.  

We present simulation results in \fig{truncate} showing focusing of metastable neon. Even when many atoms reach the substrate far from $\langle r \rangle$, the density of atoms there is orders of magnitude less than near the focus. There are two dominant mechanisms, apart from limited collimation, by which the spot is broadened: aberrations, and van der Waals attractions within the substrate and mask. Aberrations appear in the numerical solution to the fields within the mask.  Van der Waals forces are modelled by including a force term in the equations of motion that was calculated numerically and is well-approximated as being proportional to $D_2^{-4} - D_1^{-4}$, where $D_2$ is the distance to the nearest edge of the tube and $D_1$ the distance to the furthest edge. For atoms that are close enough to the wall of the tube, this force either causes the atom to collide with the tube or to strike the substrate far from the focused spot. Atoms that strike the tube walls are removed from the simulation, because they will, with high probability, either release their internal energy, making them useless for patterning, or scatter inelastically and not be focused. Although we obtain a simulated spot size of $w_0 = 8.5$~nm, the diffraction-limited spot size $w_d$ for our parameters is $w_d = 1.22 \lambda_{dB} f / (2 r_m) = 13.2$~nm, where the deBroglie wavelength $\lambda_{dB} = h/(mv) \approx 0.25$~nm. Therefore, our simulations suggest that we can focus to a diffraction-limited spot size. 


\begin{figure}
\begin{center}
\includegraphics[width=.45\textwidth]{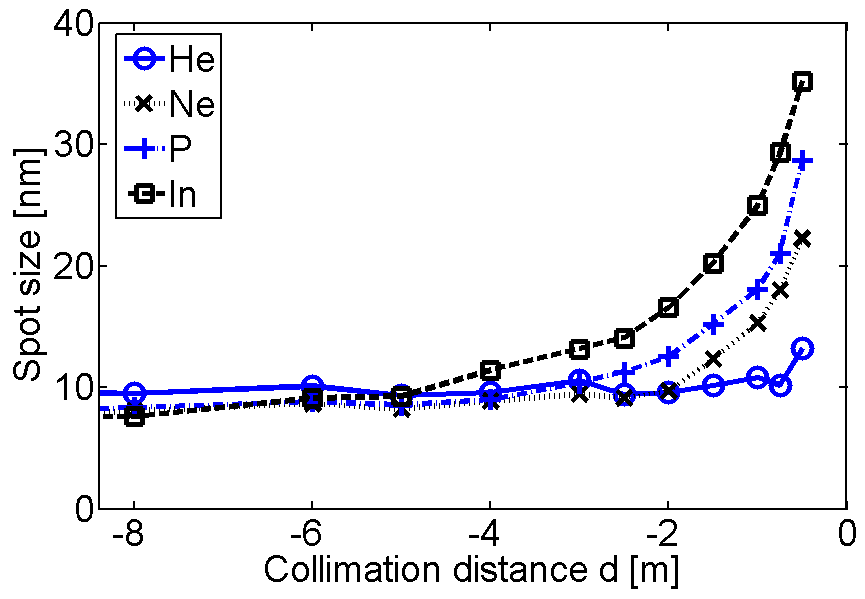}
\end{center} 
\caption{Simulated spot size $w_0$ as a function of collimation distance $d$ for four atomic species. All species except phosphorus are in a metastable excited state (given in Table~I). These large-$d$ spot sizes are smaller than the diffraction-limited spot sizes, also reported in Table~I. Helium has the largest spot size at large $d$ because it is affected more, due to its low mass, by van der Waals forces, while at small $d$, the spot is smaller due to its relatively high ratio of magnetic moment to mass. }
\label{fig:collim}
\end{figure}

Highlighting the generality of our method, our simulations show that the same mask can focus a very wide range of atomic masses with spot sizes of $\ord{10\rm{~nm}}$. \fig{collim} shows the spot sizes as a function of collimation distance $d$. For low $d$, the radial velocity spread dominates, while for high $d$, van der Waals forces and spherical aberrations dominate. In Table~I, we report the expected spot size and focal length for several species of interest at a fixed value of $d = 2$~m. For some atoms, laser excitation at a single frequency will be necessary, since they are either non-paramagnetic or have too small a magnetic moment in the ground state. For these species, a single photon will suffice to pump the atom (with some probability) into a suitable metastable state, which must have a lifetime larger than the time it takes for the atom to be deposited (typically a few milliseconds). Two notable examples are indium and gallium, which each occupy a $^{2}\mathrm{P}_{1/2}$ ground state, with maximal magnetic moment $\mu_B$/3. A single photon, at 410~nm for In and at 403~nm for Ga, would pump the atoms to a metastable $^{2}\mathrm{P}_{3/2}$ state, with maximal magnetic moment $6 \mu_B$, a state that is focused very well. The branching ratio into the desired state is 38\% for In and 67\% for Ga \footnote{Atomic data taken from the NIST Handbook of Basic Atomic Spectroscopic Data, http://www.nist.gov/physlab/data/handbook/index.cfm}. 

\begin{table}
\begin{center}
\begin{tabular*}{.45\textwidth}{@{\extracolsep{\fill}} c c c c c c} 
\hline \hline
Species &  $|\vec{\mu}|$ ($\mu_B$) & $w_0$ (nm) & $w_d$ (nm) & $f$ ($\mu$m) & State \\
\hline
$^4$He & 2 & 8.5 & 14.6 & 7.2 & $^3\mathrm{S}_1$ \\
$^{20}$Ne & 3 & 10.0 & 13.2 & 32.6 & $^3\mathrm{P}_2$ \\
$^{31}$P & 3 & 12.1 & 11.7 & 44.6 & $^4\mathrm{S}_{3/2}$ \\
$^{70}$Ga &  6 & 11.7 & 6.2 & 53.7 & $^2\mathrm{P}_{3/2}$ \\
$^{114}$In & 6 & 17.1 & 5.9 & 82.3 & $^2\mathrm{P}_{3/2}$ \\
\hline 
\end{tabular*} 
\caption{Spot sizes ($w_0$) and focal lengths ($f$) for a variety of species. All are travelling at 400~m/s and have traversed 2~m following a 250~$\mu$m skimmer. The approximate diffraction-limited spot size $w_d$ is also given; the actual spot size is expected to be the convolution of the atomic distribution of waist $w_0$ and the Airy disc of diameter $w_d$.  The magnetic moment ($|\vec{\mu}|$), and atomic state are given for reference. All except $^{31}$P are in metastable states. The width of both the substrate and the mask is 150~$\mu$m.}
\label{fig:eltstable}
\end{center} 
\end{table}

For several of the species we examined, the spot size is limited by diffraction from the mask aperture.  Since the diameter of the Airy disc is given by $w_d = 1.22 \lambda_{dB} f / (2 r_m)$, a decrease in the ratio ${f}/{r_m}$ will reduce the diffraction limit.  Simulations with multiple layers of thin film magnetic material show a decreased focal length with identical mask apertures, resulting in a smaller diffraction limit with similar or better spot sizes.  Alternatively, advances in the science of thin film magnetic materials may provide for greater magnetizations, which would similarly reduce the focal length.  As such, we do not regard the calculated diffraction limits for our simulated apparatus as being the best that can be achieved with this method. 


Among the many potential applications of our method, one of the most intriguing is the fabrication of quantum dots. Currently, quantum dots are most frequently fabricated by molecular beam epitaxy, which results in quantum dots of random size and location. Our scheme could produce dots with a position known to within a few nm and a size limited only by the Poisson fluctuations. Mounting the mask on a nanometer-resolution translation stage increases the versatility of our method. For instance, the quantum dots could be combined with nanofabricated wires or mirrors, producing electrical or optical interconnects between the dots. This could enable one vision of quantum computation with quantum dots \cite{Loss:98,Imamoglu:99}. Our method has many other potential uses, as well:  applications to basic science include plasmonics \cite{Schuller:10}, metamaterials \cite{Shalaev:07}, and quantum photonics \cite{Obrien:09}, while applications of commercial interest include photovoltaics, light sources, and light sensors. 

In conclusion, we have outlined a new method for fabricating a wide variety of nanoscale devices with an unprecedented combination of nanometer precision and high throughput. Our method relies on technologies that are well understood, including supersonic beams and magnetic filtering, and has no strenuous laser requirements. One element that remains to be developed is the magnetic mask, but the required magnetic film and e-beam patterning technology are readily available. Our method should open up new possibilities in the fabrication of nanoscale semiconductor quantum devices, including light sources and detectors, plasmonic devices, and quantum information processors. 

We gratefully acknowledge fruitful discussions with Dan Ralph, David Sellmyer, Andrew Houck, Walt de Heer, and Rene Gerritsma, and funding from the Sid W. Richardson Foundation.


\begin{thebibliography}{33}
\expandafter\ifx\csname natexlab\endcsname\relax\def\natexlab#1{#1}\fi
\expandafter\ifx\csname bibnamefont\endcsname\relax
  \def\bibnamefont#1{#1}\fi
\expandafter\ifx\csname bibfnamefont\endcsname\relax
  \def\bibfnamefont#1{#1}\fi
\expandafter\ifx\csname citenamefont\endcsname\relax
  \def\citenamefont#1{#1}\fi
\expandafter\ifx\csname url\endcsname\relax
  \def\url#1{\texttt{#1}}\fi
\expandafter\ifx\csname urlprefix\endcsname\relax\def\urlprefix{URL }\fi
\providecommand{\bibinfo}[2]{#2}
\providecommand{\eprint}[2][]{\url{#2}}

\bibitem[{\citenamefont{Rai-Choudhury}(1997)}]{microfab:book}
\bibinfo{editor}{\bibfnamefont{P.}~\bibnamefont{Rai-Choudhury}}, ed.,
  \emph{\bibinfo{title}{Handbook of Microlithography, Micromachining, and
  Microfabrication}} (\bibinfo{publisher}{SPIE Optical Engineering Press and
  the Institution of Electrical Engineers}, \bibinfo{year}{1997}).

\bibitem[{\citenamefont{O'Brien et~al.}(2009)\citenamefont{O'Brien, Furusawa,
  and Vuckovic}}]{Obrien:09}
\bibinfo{author}{\bibfnamefont{J.~L.} \bibnamefont{O'Brien}},
  \bibinfo{author}{\bibfnamefont{A.}~\bibnamefont{Furusawa}}, \bibnamefont{and}
  \bibinfo{author}{\bibfnamefont{J.}~\bibnamefont{Vuckovic}},
  \bibinfo{journal}{Nature Photonics} \textbf{\bibinfo{volume}{3}},
  \bibinfo{pages}{687} (\bibinfo{year}{2009}).

\bibitem[{\citenamefont{Dubertret et~al.}(2002)\citenamefont{Dubertret,
  Skourides, Norris, Noireaux, Brivanlou, and Libchaber}}]{Dubertret:02}
\bibinfo{author}{\bibfnamefont{B.}~\bibnamefont{Dubertret}} \bibnamefont{\emph{et~al.}},
  \bibinfo{journal}{Science} \textbf{\bibinfo{volume}{298}},
  \bibinfo{pages}{1759} (\bibinfo{year}{2002}).

\bibitem[{\citenamefont{Chen and P\'{e}pin}(2001)}]{Chen:01}
\bibinfo{author}{\bibfnamefont{Y.}~\bibnamefont{Chen}} \bibnamefont{and}
  \bibinfo{author}{\bibfnamefont{A.}~\bibnamefont{P\'{e}pin}},
  \bibinfo{journal}{Electrophoresis} \textbf{\bibinfo{volume}{22}},
  \bibinfo{pages}{187} (\bibinfo{year}{2001}).

\bibitem[{\citenamefont{Orloff}(1993)}]{Orloff:93}
\bibinfo{author}{\bibfnamefont{J.}~\bibnamefont{Orloff}},
  \bibinfo{journal}{Rev. Sci. Instr.} \textbf{\bibinfo{volume}{64}},
  \bibinfo{pages}{1105} (\bibinfo{year}{1993}).

\bibitem[{\citenamefont{Meschede and Metcalf}(2003)}]{Meschede:03}
\bibinfo{author}{\bibfnamefont{D.}~\bibnamefont{Meschede}} \bibnamefont{and}
  \bibinfo{author}{\bibfnamefont{H.}~\bibnamefont{Metcalf}},
  \bibinfo{journal}{J. Phys. D.} \textbf{\bibinfo{volume}{36}},
  \bibinfo{pages}{R17} (\bibinfo{year}{2003}).

\bibitem[{\citenamefont{Balykin and Melent'ev}(2009)}]{Balykin:09b}
\bibinfo{author}{\bibfnamefont{V.~I.} \bibnamefont{Balykin}} \bibnamefont{and}
  \bibinfo{author}{\bibfnamefont{P.~N.} \bibnamefont{Melent'ev}},
  \bibinfo{journal}{Nanotechnologies in Russia} \textbf{\bibinfo{volume}{4}},
  \bibinfo{pages}{425} (\bibinfo{year}{2009}).

\bibitem[{\citenamefont{Johnson et~al.}(1998)\citenamefont{Johnson, Thywissen,
  Dekker, Berggren, Chu, Younkin, and Prentiss}}]{Johnson:98}
\bibinfo{author}{\bibfnamefont{K.~S.} \bibnamefont{Johnson}} \bibnamefont{\emph{et~al.}},
  \bibinfo{journal}{Science} \textbf{\bibinfo{volume}{280}},
  \bibinfo{pages}{1583} (\bibinfo{year}{1998}).

\bibitem[{\citenamefont{Baker et~al.}(2004)\citenamefont{Baker, Palmer,
  MacGillivray, and Sang}}]{Baker:04}
\bibinfo{author}{\bibfnamefont{M.}~\bibnamefont{Baker}},
  \bibinfo{author}{\bibfnamefont{A.~J.} \bibnamefont{Palmer}},
  \bibinfo{author}{\bibfnamefont{W.~R.} \bibnamefont{MacGillivray}},
  \bibnamefont{and} \bibinfo{author}{\bibfnamefont{R.~T.} \bibnamefont{Sang}},
  \bibinfo{journal}{Nanotechnology} \textbf{\bibinfo{volume}{15}},
  \bibinfo{pages}{1356} (\bibinfo{year}{2004}).

\bibitem[{\citenamefont{Lu et~al.}(1998)\citenamefont{Lu, Baldwin, Hoogerland,
  Buckman, Senden, Sheridan, and Boswell}}]{Lu:98}
\bibinfo{author}{\bibfnamefont{W.}~\bibnamefont{Lu}}, \bibnamefont{\emph{et~al.}},
  \bibinfo{journal}{J. Vac. Sci. Technol. B}
  \textbf{\bibinfo{volume}{16}}, \bibinfo{pages}{3846} (\bibinfo{year}{1998}).

\bibitem[{\citenamefont{Bard et~al.}(1997)\citenamefont{Bard, Berggren, Wilbur,
  Gillaspy, Rolston, McClelland, Phillips, Prentiss, and Whitesides}}]{Bard:97}
\bibinfo{author}{\bibfnamefont{A.}~\bibnamefont{Bard}} \bibnamefont{\emph{et~al.}},
  \bibinfo{journal}{J. Vac. Sci. Technol. B} \textbf{\bibinfo{volume}{15}},
  \bibinfo{pages}{1805} (\bibinfo{year}{1997}).

\bibitem[{\citenamefont{Engels et~al.}(1999)\citenamefont{Engels, Salewski,
  Levsen, Sengstock, and Ertmer}}]{Engels:99}
\bibinfo{author}{\bibfnamefont{P.}~\bibnamefont{Engels}},
  \bibinfo{author}{\bibfnamefont{S.}~\bibnamefont{Salewski}},
  \bibinfo{author}{\bibfnamefont{H.}~\bibnamefont{Levsen}},
  \bibinfo{author}{\bibfnamefont{K.}~\bibnamefont{Sengstock}},
  \bibnamefont{and} \bibinfo{author}{\bibfnamefont{W.}~\bibnamefont{Ertmer}},
  \bibinfo{journal}{Appl. Phys. B} \textbf{\bibinfo{volume}{69}},
  \bibinfo{pages}{407} (\bibinfo{year}{1999}).

\bibitem[{\citenamefont{McClelland et~al.}(1993)\citenamefont{McClelland,
  Scholten, Palm, and Celotta}}]{McClelland:93}
\bibinfo{author}{\bibfnamefont{J.~J.} \bibnamefont{McClelland}},
  \bibinfo{author}{\bibfnamefont{R.~E.} \bibnamefont{Scholten}},
  \bibinfo{author}{\bibfnamefont{E.~C.} \bibnamefont{Palm}}, \bibnamefont{and}
  \bibinfo{author}{\bibfnamefont{R.~J.} \bibnamefont{Celotta}},
  \bibinfo{journal}{Science} \textbf{\bibinfo{volume}{262}},
  \bibinfo{pages}{877} (\bibinfo{year}{1993}).

\bibitem[{\citenamefont{Timp et~al.}(1992)\citenamefont{Timp, Behringer,
  Tennant, Cunningham, Prentiss, and Berggren}}]{Timp:92}
\bibinfo{author}{\bibfnamefont{G.}~\bibnamefont{Timp}} \bibnamefont{\emph{et~al.}},
  \bibinfo{journal}{Phys. Rev. Lett.} \textbf{\bibinfo{volume}{69}},
  \bibinfo{pages}{1636} (\bibinfo{year}{1992}).

\bibitem[{\citenamefont{Natarajan et~al.}(1996)\citenamefont{Natarajan,
  Behringer, and Timp}}]{Natarajan:96}
\bibinfo{author}{\bibfnamefont{V.}~\bibnamefont{Natarajan}},
  \bibinfo{author}{\bibfnamefont{R.~E.} \bibnamefont{Behringer}},
  \bibnamefont{and} \bibinfo{author}{\bibfnamefont{G.}~\bibnamefont{Timp}},
  \bibinfo{journal}{Phys. Rev. A} \textbf{\bibinfo{volume}{53}},
  \bibinfo{pages}{4381} (\bibinfo{year}{1996}).

\bibitem[{\citenamefont{Gupta et~al.}(1995)\citenamefont{Gupta, McClelland,
  Jabbour, and Celotta}}]{Gupta:95}
\bibinfo{author}{\bibfnamefont{R.}~\bibnamefont{Gupta}},
  \bibinfo{author}{\bibfnamefont{J.~J.} \bibnamefont{McClelland}},
  \bibinfo{author}{\bibfnamefont{Z.~J.} \bibnamefont{Jabbour}},
  \bibnamefont{and} \bibinfo{author}{\bibfnamefont{R.~J.}
  \bibnamefont{Celotta}}, \bibinfo{journal}{Appl. Phys. Lett.}
  \textbf{\bibinfo{volume}{67}}, \bibinfo{pages}{1378} (\bibinfo{year}{1995}).

\bibitem[{\citenamefont{Smeets et~al.}(2009)\citenamefont{Smeets, van~der
  Staten, Meijer, Fabrie, and van Leeuwen}}]{Smeets:09}
\bibinfo{author}{\bibfnamefont{B.}~\bibnamefont{Smeets}},
  \bibinfo{author}{\bibfnamefont{P.}~\bibnamefont{van~der Staten}},
  \bibinfo{author}{\bibfnamefont{T.}~\bibnamefont{Meijer}},
  \bibinfo{author}{\bibfnamefont{C.~G. C. H.~M.} \bibnamefont{Fabrie}},
  \bibnamefont{and} \bibinfo{author}{\bibfnamefont{K.~A.~H.} \bibnamefont{van
  Leeuwen}}, \bibinfo{journal}{Appl. Phys. B} \textbf{\bibinfo{volume}{98}},
  \bibinfo{pages}{697} (\bibinfo{year}{2009}).

\bibitem[{\citenamefont{Berggren et~al.}(1995)\citenamefont{Berggren, Bard,
  Wilbur, Gillaspy, Helg, McClelland, Rolston, Phillips, Prentiss, and
  Whitesides}}]{Berggren:95}
\bibinfo{author}{\bibfnamefont{K.~K.} \bibnamefont{Berggren}} \bibnamefont{\emph{et~al.}},
\bibinfo{journal}{Science}
  \textbf{\bibinfo{volume}{269}}, \bibinfo{pages}{1255} (\bibinfo{year}{1995}).

\bibitem[{\citenamefont{Averbukh and Arvieu}(2001)}]{Averbukh:01}
\bibinfo{author}{\bibfnamefont{I.~S.} \bibnamefont{Averbukh}} \bibnamefont{and}
  \bibinfo{author}{\bibfnamefont{R.}~\bibnamefont{Arvieu}},
  \bibinfo{journal}{Phys. Rev. Lett.} \textbf{\bibinfo{volume}{87}},
  \bibinfo{pages}{163601} (\bibinfo{year}{2001}).

\bibitem[{\citenamefont{Oskay et~al.}(2002)\citenamefont{Oskay, Steck, and
  Raizen}}]{Oskay:02}
\bibinfo{author}{\bibfnamefont{W.~H.} \bibnamefont{Oskay}},
  \bibinfo{author}{\bibfnamefont{D.~A.} \bibnamefont{Steck}}, \bibnamefont{and}
  \bibinfo{author}{\bibfnamefont{M.~G.} \bibnamefont{Raizen}},
  \bibinfo{journal}{Phys. Rev. Lett.} \textbf{\bibinfo{volume}{89}},
  \bibinfo{pages}{283001} (\bibinfo{year}{2002}).

\bibitem[{\citenamefont{K\"{a}nders et~al.}(1995)\citenamefont{K\"{a}nders,
  Lison, Richter, Wynands, and Meschede}}]{Kaenders:95}
\bibinfo{author}{\bibfnamefont{W.~G.} \bibnamefont{K\"{a}nders}},
  \bibinfo{author}{\bibfnamefont{F.}~\bibnamefont{Lison}},
  \bibinfo{author}{\bibfnamefont{A.}~\bibnamefont{Richter}},
  \bibinfo{author}{\bibfnamefont{R.}~\bibnamefont{Wynands}}, \bibnamefont{and}
  \bibinfo{author}{\bibfnamefont{D.}~\bibnamefont{Meschede}},
  \bibinfo{journal}{Nature} \textbf{\bibinfo{volume}{375}},
  \bibinfo{pages}{214} (\bibinfo{year}{1995}).

\bibitem[{\citenamefont{Melentiev et~al.}(2009)\citenamefont{Melentiev,
  Zablotskiy, Lapshin, Sheshin, Baturin, and Balykin}}]{Balykin:09c}
\bibinfo{author}{\bibfnamefont{P.~N.} \bibnamefont{Melentiev}},
  \bibinfo{author}{\bibfnamefont{A.~V.} \bibnamefont{Zablotskiy}},
  \bibinfo{author}{\bibfnamefont{D.~A.} \bibnamefont{Lapshin}},
  \bibinfo{author}{\bibfnamefont{E.~P.} \bibnamefont{Sheshin}},
  \bibinfo{author}{\bibfnamefont{A.~S.} \bibnamefont{Baturin}},
  \bibnamefont{and} \bibinfo{author}{\bibfnamefont{V.~I.}
  \bibnamefont{Balykin}}, \bibinfo{journal}{Nanotechnology}
  \textbf{\bibinfo{volume}{20}}, \bibinfo{pages}{235301}
  (\bibinfo{year}{2009}).

\bibitem[{\citenamefont{Campargue}(2001)}]{Campargue:01}
\bibinfo{editor}{\bibfnamefont{R.}~\bibnamefont{Campargue}}, ed.,
  \emph{\bibinfo{title}{Atom and Molecular Beams: The State of the Art 2000}}
  (\bibinfo{publisher}{Springer-Verlag, Berlin}, \bibinfo{year}{2001}).

\bibitem[{\citenamefont{Pauly}(2000)}]{Pauly:00}
\bibinfo{author}{\bibfnamefont{H.}~\bibnamefont{Pauly}},
  \emph{\bibinfo{title}{Atom, Molecule and Clusterbeams I: Basic Theory,
  Production, and Detection of Thermal Energy Beams}}
  (\bibinfo{publisher}{Springer-Verlag, Berlin}, \bibinfo{year}{2000}).

\bibitem[{\citenamefont{Buckland}(1998)}]{Buckland:thesis}
\bibinfo{author}{\bibfnamefont{J.~R.} \bibnamefont{Buckland}}, Ph.D. thesis,
  \bibinfo{school}{University of Cambridge}, \bibinfo{address}{Cambridge, UK}
  (\bibinfo{year}{1998}).

\bibitem[{\citenamefont{Hoogerland et~al.}(1994)\citenamefont{Hoogerland,
  Driessen, Vredenbregt, Megens, Schuwer, Beijerinck, and van
  Leeuwen}}]{Hoogerland:94}
\bibinfo{author}{\bibfnamefont{M.~D.} \bibnamefont{Hoogerland}} \bibnamefont{\emph{et~al.}},
  in \emph{\bibinfo{booktitle}{Proceedings of the 1994 IEEE
  International Frequency Control Symposium}} (\bibinfo{publisher}{IEEE},
  \bibinfo{year}{1994}), p. \bibinfo{pages}{651}.

\bibitem[{\citenamefont{Rasel et~al.}(1999)\citenamefont{Rasel, Santos, Pavone,
  Perales, Unnikrishnan, and Leduc}}]{Rasel:99}
\bibinfo{author}{\bibfnamefont{E.}~\bibnamefont{Rasel}},
  \bibinfo{author}{\bibfnamefont{F.~P.~D.} \bibnamefont{Santos}},
  \bibinfo{author}{\bibfnamefont{F.~S.} \bibnamefont{Pavone}},
  \bibinfo{author}{\bibfnamefont{F.}~\bibnamefont{Perales}},
  \bibinfo{author}{\bibfnamefont{C.~S.} \bibnamefont{Unnikrishnan}},
  \bibnamefont{and} \bibinfo{author}{\bibfnamefont{M.}~\bibnamefont{Leduc}},
  \bibinfo{journal}{Eur. J. Phys. D.} \textbf{\bibinfo{volume}{7}},
  \bibinfo{pages}{311} (\bibinfo{year}{1999}).

\bibitem[{\citenamefont{Xing et~al.}(2007)\citenamefont{Xing, Barb, Gerritsma,
  Spreeuw, Luigjes, Xiao, R'{e}tif, and Goedkoop}}]{Xing:07}
\bibinfo{author}{\bibfnamefont{Y.~T.} \bibnamefont{Xing}} \bibnamefont{\emph{et~al.}},
  \bibinfo{journal}{J. Magn. Magn. Mater.} \textbf{\bibinfo{volume}{313}},
  \bibinfo{pages}{192} (\bibinfo{year}{2007}).

\bibitem[{\citenamefont{Fernholz et~al.}(2008)\citenamefont{Fernholz,
  Gerritsma, Whitlock, Barb, and Spreeuw}}]{Fernholz:08}
\bibinfo{author}{\bibfnamefont{T.}~\bibnamefont{Fernholz}},
  \bibinfo{author}{\bibfnamefont{R.}~\bibnamefont{Gerritsma}},
  \bibinfo{author}{\bibfnamefont{S.}~\bibnamefont{Whitlock}},
  \bibinfo{author}{\bibfnamefont{I.}~\bibnamefont{Barb}}, \bibnamefont{and}
  \bibinfo{author}{\bibfnamefont{R.~J.~C.} \bibnamefont{Spreeuw}},
  \bibinfo{journal}{Phys. Rev. A} \textbf{\bibinfo{volume}{77}},
  \bibinfo{pages}{033409} (\bibinfo{year}{2008}).

\bibitem[{\citenamefont{Loss and DiVincenzo}(1998)}]{Loss:98}
\bibinfo{author}{\bibfnamefont{D.}~\bibnamefont{Loss}} \bibnamefont{and}
  \bibinfo{author}{\bibfnamefont{D.~P.} \bibnamefont{DiVincenzo}},
  \bibinfo{journal}{Phys. Rev. A} \textbf{\bibinfo{volume}{57}},
  \bibinfo{pages}{120} (\bibinfo{year}{1998}).

\bibitem[{\citenamefont{Imamoglu et~al.}(1999)\citenamefont{Imamoglu,
  Awschalom, Burkard, DiVincenzo, Loss, Sherwin, and Small}}]{Imamoglu:99}
\bibinfo{author}{\bibfnamefont{A.}~\bibnamefont{Imamoglu}} \bibnamefont{\emph{et~al.}},
  \bibinfo{journal}{Phys. Rev. Lett.} \textbf{\bibinfo{volume}{83}},
  \bibinfo{pages}{4204} (\bibinfo{year}{1999}).

\bibitem[{\citenamefont{Schuller et~al.}(2010)\citenamefont{Schuller, Barnard,
  Cai, Jun, White, and Brongersma}}]{Schuller:10}
\bibinfo{author}{\bibfnamefont{J.~A.} \bibnamefont{Schuller}} \bibnamefont{\emph{et~al.}},
  \bibinfo{journal}{Nature Materials} \textbf{\bibinfo{volume}{9}},
  \bibinfo{pages}{193} (\bibinfo{year}{2010}).

\bibitem[{\citenamefont{Shalaev}(2007)}]{Shalaev:07}
\bibinfo{author}{\bibfnamefont{V.~M.} \bibnamefont{Shalaev}},
  \bibinfo{journal}{Nature Photonics} \textbf{\bibinfo{volume}{1}},
  \bibinfo{pages}{41} (\bibinfo{year}{2007}).

\end{thebibliography}

\bibliographystyle{apsrev}

\end{document}